# Phishing Attacks: Detection And Prevention

Mario Ciprian Birlea
Faculty of Engineering and Informatics
University of Bradford
Bradford, UK
M.C.Birlea@bradford.ac.uk

*Abstract*—This paper aims to provide an understanding of what a phishing attack is, the types of phishing attacks and methods employed by cyber criminals. This journal will also provide an understanding on where phishing attacks are carried via various technologies and provide an understanding of working actively to detect and proactively to prevent. As we are continuously developing the technology the chance of cyber-attacks will proportionally increase. Consequently, this paper will help the readers understand what studies have been conducted, analyzed and results provided pinpointed the essence of phishing attacks.

## I. Introduction

Phishing is a type of identity theft when a cybercriminal impersonates another person or organisation with the purpose of retrieving through illegal purposes sensitive information such as account details, login password, credit card numbers details. Phishing has been shown that originated somewhere around 1995, over 24 years ago stated by AOL [2,3]. The term of phishing is an alternative of the term 'fishing' and in a similar way the attacker is launching a 'bait' to the victim and tricks it while gathering the information required.

Over the past years the AWPG from 2017 has published that the phishing attacks have increased over a period of 12 years with an average monthly increase of 5753% [2]. The authorities reported to engage and educate the general public. To better understand the way a cyber criminal operates there is a need to take a closer look at understanding different methods, detecting this anomaly and understanding the types and preventions of this malicious activity [6].

## II. Insinde Cyber criminal's Mind

This type of cyber-attack started as a social engineering scam back in 1995. The cyber criminal is impersonating a company or a specific person targeting sensitive information that has a potential to bring financial gain and sensitive information. It has been identified that one of the primary channels when this method has occurred has been seen through instant messaging (IMs) clients that were very popular through 1990's such as Yahoo, MSM. To better understand the methods employed by cybercriminals below it is being described what are the phishing stages performed before, during and after an attack. Are the steps presented below are equally important and interdependent for the attack to be carried successfully. Furthermore, Phishing attacks might be the step to get the point of entry in Advanced Persistent Threat (APT) [14, 16].

### A. Attack preparation & planning

During this initial step the attacker decides on the best media channel to attack its victim. Even though the instant messaging has been as the primary channel in 1990s the most common communication channel is email but not only. There are social media applications, phone applications and many others.

In addition, after selection of the channel, the attacker is selecting the device, the attack technique and starts to prepare the necessary materials to carry the attack. This attack can be performed manually or automated tools. Kali Linux has incorporated a 'social engineering toolkit'. This tool empowers the user to clone a specific website such as: Facebook, Yahoo, Google and others. In the screenshot below Kali social engineering toolkit is being presented along with the 1st step and cloning technique used for Facebook, social media webpage. The method presented represents only one of the possible methods a cybercriminal could use to prepare its materials and clone a social medial website.

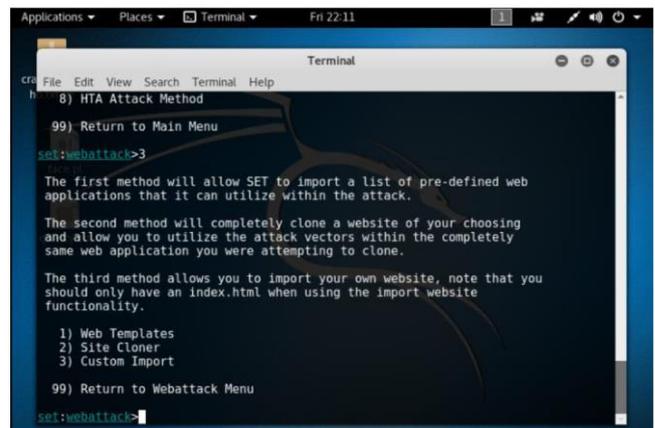

*Figure 1 – Cloning a website using Kali Linux – Social Engineering tool [13]*

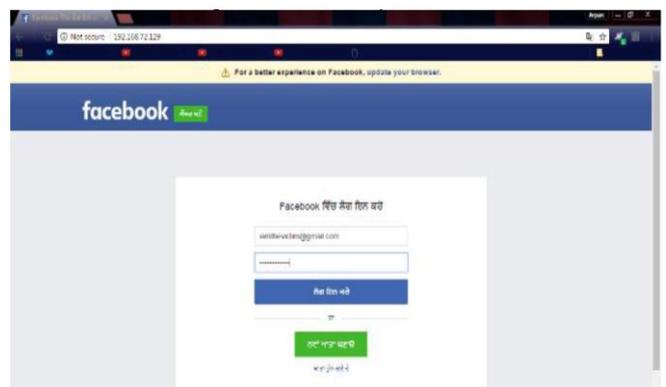

*Figure 2 – Website clone results. Phishing attacks and its countermeasures [13]*

### B. Attack execution

This second step in the process consist of three other sub-steps such as:

- *Attack material distribution*. Attacker considers the amount of people he has targeted and their devices

considering the end scope. This factor of materials distribution heavily relies on the user's device type [17].
- *Target data collection.* This process starts as soon as the victim responds to the phisher's materials.
- *Target resource penetration.* Attackers may compromise system by injecting a website with client-side script to collect user information. This last step is optional in the process as it depends on the end scope of the attack and the way the cybercriminal operates [8].

### C. Attack results explotation

The last step of the attack after acquiring victims desire information are usually to impersonate them, therefore closing the phishing lifecycle [17]. Please find below illustrated in *Figure 3- Lifecycle of the phishing attack* the lifecycle of phishing attacks.

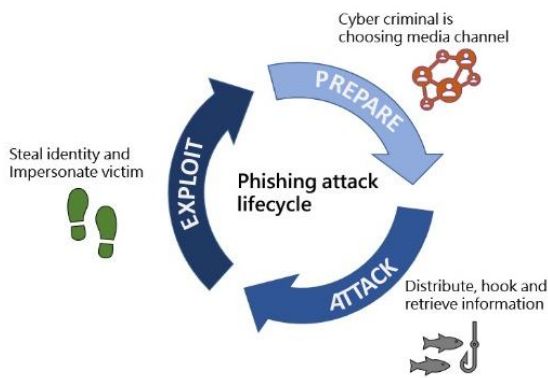

*Figure 3 – Lifecycle of the phishing attack*

### III. TYPES OF PSHISHING ATTACKS

The term of "detection" is often associated with the action of a detective, to investigate a crime with the purpose of finding the responsible person. Like the detective, using the right tools and specific steps could lead to mitigation of the problem. It is important to differentiate the attack types and correct process to uncover the phishing attack [12, 20].

According to a survey from 2 years ago [13] around 10.2% of the Phishing emails get through the protection software and 5% of those are being opened. Even though that numbers sounds somewhat small, the truth is that out of the 156 million phishing emails, 8 million infected mails are opened. This can be compared to the size of the capital of United Kingdom. If we consider the identify theft and economic opportunities, it gives us a clearer view. Therefore, let's investigate what type of attacks are, how to detect and prevent them (please see section IV). Besides, the compromised machines can be used as part of a botnet [11, 18].

### A. Deceptive Phising

This method if often used by cyber criminals to impersonate a legitimate company and steal victim login credentials. After the attacker has been successful, he will use the information he could steal after impersonating the user as leverage and will manipulate/blackmail to obtain further gain.

### B. Spear Phishing

The FBI has highlighted that the art of deception and the criminal organisations have employed more sophisticated methods. Spear-phishing being one of the 3 top online tools used by cybercriminal masterminds and do billions of dollars and affects all organisations from well-known corporations to non-profit, even churches and schools [13]. To understand spear phishing attack, there are three questions to have in mind: what is it? What software? What opportunity?

- Spear finishing is a cyber attack carried via suspicious activities targeting an individual
- There is no such a thing as a *perfect protection software*. Wireless IDPS (intrusion detection prevention system) is analysing the traffic of the network protocol activities, detects unauthorised LAN (local area network) is use [6].
- Finally, this gives the cybercriminal the opportunity to find vulnerabilities and infiltrate in the application layer, transport or protocol activities.

### C. Whaling

The term of "whaling" refers to the victim as being highly ranked or important like comparing the size and importance to the whale, as a trophy. Like spear phishing, whaling method targets a specific individual that is a very powerful person, or a very wealthy person. Here, the main scope for the attacker is to gain control over different accounts of the victim and especially expose sensitive information to social media or black mail the victim for different gains [6].

### D. Phishing by malwayre software

The major organisations that manipulate very sensitive user data (bank cards, personal details etc.) are usually investing highly in different software applications to protect themselves against malware attacks. Although this process is efficient to some extent, there is a human component that cannot be neglected. Phishing also called "social engineering" vis the use of technology and it has been proven many times to be very efficient.

Early in 2016, Target organisation (USA) was the victim of major data breach. The cyber criminals behind this attack, targeted a trusted machine within Target infrastructure, instead of directly attacking the system. After stealing credit card details, they remove their tracks by bouncing back this attack on different malware infected machines around the globe [13].

Another example of this types of attack can occur with providing malicious attachments such as a "keylogger" to collect user credentials and expose them to the hacker. To prevent the keyloggers there are different software we can use such as Key Scrambler and Spy Shelter [13].

### E. Search engine phishing

Another example of phishing attack is search engine phishing. Through this method, cybercriminals are creating appealing websites that offer products or services under the market price and lure the users to purchase them while appearing to legal. The reality is that websites are fake, and their creator is taking the user details and uses them for malicious purposes [13].

A good example can be provided by a hotel booking website, this can appear genuine but during the payment the

"https:/" will not appear and the padlock icon would be missing [15]. Those are only 2 giveaways that the website is not genuine. What will happen next is the cybercriminals could clone the card details make different payments and quickly clearing the funds. In 2015 more than 2.5 million users have been impacted, please Figure 4 from below for more details.

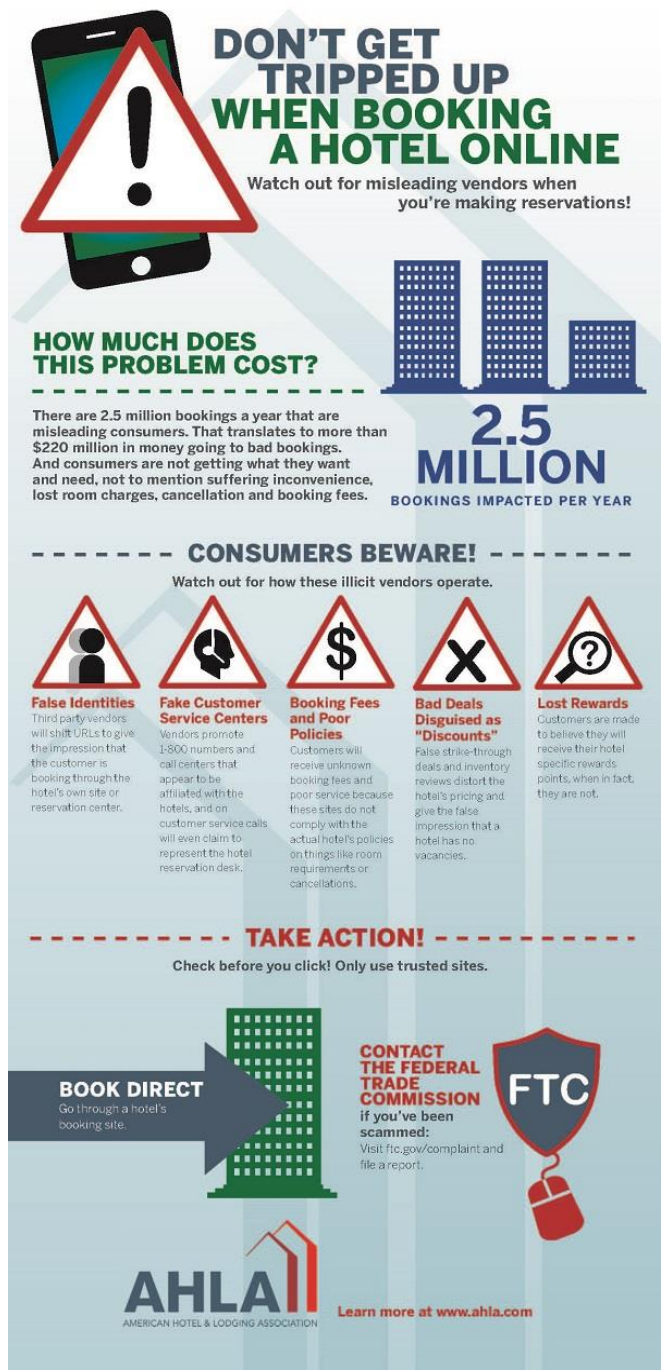

*Figure 4 – Search engine phishing, hotel scams [15]*

### F. Man-in-the-middle phishing

Man-in-the-middle phishing attack is defined when the cybercriminal places himself between a user and a legal website or system. The framework of their method is that the hacker is reading the traffic data in between (email, card details, private information etc.), filters and saves what he considers a leverage [13].

A good example can be provided of a hacker that identified ongoing conversation between a remote employee and organisation. The HR department could request for the updated credit card details via email from another employee. In this scenario the cyber criminal can stop the traffic, update the email before and release to HR with another credit card details.

### G. Voice phishing

The cyber criminals could use their social engineering skills via the telephone system to obtain private and financial information [19].

Some of the cyber criminals try to fraud the victims mentioning scams. A car accident would be a good example. The attacker tries to pretend that a family member of the victim was involved in a car crash and they can claim a very high compensation. During this call the cybercriminal is asking for the credit details and/or personal information. The government of Australia created a specific page and section with more details to support victims of the attack [17].

## IV. DETECTION AND PREVENTION OF PHISHING

The section presented above a detailed view of the different types of phishing attacks with relevant examples. In this section there are presented detection methods and advice to prevent these specific attacks.

### A. Phishing emails

The attackers could use various phishing methods incorporated in the email (i.e. deceptive, spear phishing, whaling). To be able to recognise a fake email as an attempt of phishing, it is critical to look at the details such as:

- Email address: this needs to be correctly spelled and concise. A possible giveaway is an email where number '0' is used instead of letter 'o'

- Greetings: the big organisations will rarely address their customers with a generic 'Dear Customer'

- Grammar mistakes: proofread the whole email text and check for spelling mistakes, there is a small chance that emails received with high importance containing typos

- Asking for urgent action: some of the cybercriminals that work on stealing information alert their victims that their account will be disabled.

Considering the above the best prevention method is to proofread the emails that we are receiving and especially high importance and the ones asking for personal details.

Do not click or download attachments from unknown senders and if they look suspicious report the specific emails. If you receive suspicions emails with requests that do not match company policy rules, report the email [19].

Big technology companies provide their employees user training for have special teams assigned to carry phishing attacks and assure their employees take the appropriate actions, this way they tackle this working both actively and proactively.

## B. Social Engineering

Cybercriminals try to trick their victims in any ways possible and this method brings the combination of tenacity, inter-personal skills and technology. The social engineering technique could be carried via any technology channel (email, phone, websites, media etc.). This is where usually the attackers impersonate someone such as a bank employee, manager or even relative [19].

In all possible environments, thus it is advised that personal details could be provided via phone, there is a need to be assertive and if there is any suspicion regarding the sender, an investigation should be carried out (mobile number, email address or website). The social engineering attacks carried via web are hard to detect as your machine will not notify you that it has been compromised, but it is good to keep an eye for website warning before providing any personal information [5].

As mentioned in III (G) the government would usually identify and provide information on scams such as car accident claim calls performed by cybercriminals. They government of Australia advises citizens to ask for caller name, organisation details, department, carry individual investigation such as web search with blacklisted phone numbers and empower them to call a government support line number for further advice [17].

## C. Man-in-the-middle prevention

This type of attack as understood above (III – F), the cybercriminal is positioning between two parties by DNS spoofing, ARP positioning or even email phishing. There are multiple organisations that lost millions of dollars due to this phishing method and as they had no clue their communication via network are being intercepted [13]. Table 1 illustrates three prevention methods.

## D. Security awarneess and training

All organisations should have in place a basic and regular training. This should provide an understanding of what a good email and phishing attacks looks like, manage them and provide prevention methods for different types of attack. The overall benefit of the proactive action is to avoid passive situation, reduce success of attacks and ensure security and management will respond accordingly [13].

## V. IMPACT ON ORGANISATIONS

After numerous attacks have been carried over the year, organisations from various sensitive information via data breaches.

1) A complex survey has been developed by Lee Neely (2017) part of SANS Institute. This survey has been sponsored by different software and anti-virus firms (Cylance, FireEye, McAfee, Qualys) and involved over 263 IT professionals. Moreover, it has been remarkable that organisations with all sizes (100-100,000 employees) from various industries have participated in this survey and provided their input. Figure 5 – SANS participants by industry, provides a more detailed look of what were the major participants.

The results from the survey has pinpointed that almost a quarter (74%) of the participants in the survey have chosen the email attachment or link to dominate the entry point of cyber threats [19]. *Figure 6 – Top directions of entry [19]* provides a more detailed look of the following breaches directions.

| Technology used | How does it work? | Software used |
|---|---|---|
| VPN | • Broadens the private netwrok across public network<br>• Hides the IP address from public websites<br>• Protects data and creates encrypted connection | Depends on the organization provider |
| Proxy Server with Data encryption | • Encrypts the transmission using a realiable and secure proxy server | OpenVPN, Tor Browser, I2P Hide my IP |
| Secure Shell Tunelling | • Logging into a remote machine and execute commands<br>• A tunnul is being created and data encrypted via secure shell protocol<br>• 3 cryptographic techniques used are: symetrical, asimetrical and hashing<br>• Supports TCP/X11 | WinSCP |

*Table 1 – Man-in-the-middle prevention [13]*

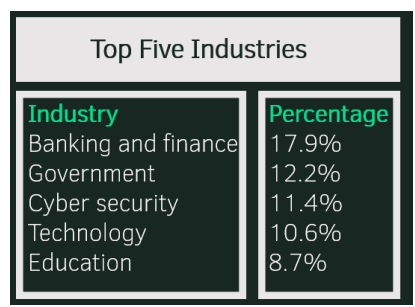

*Figure 5 – SANS participants by industry [19]*

Another look at the granularity of the data has shown that phishing, spear phishing, and whaling were the top threats entering in the organisations with an impressive 40%. In addition, it has been reported that the top" malware-less" threats are credential compromise (22%) [19].

2) Karsperky released a publication in 2017 providing a deeper view into the malware and other security threats. This article is providing details of the worring increase of the dominant finacial threat – phishing. They have detected an increase of almost 19% from 2014 to 2016 in financial phishing attacks. Furthemore the finance and banking indutry are posing a risk as finanicla pshishing attacks are increasing to an average rate of approx. 9.37% year by year [21]. Please reffer to Figure 7 – Percentage of financial phishing 2014-2016 [21] for more details. The aricle covers in detail malware and also andoid attacks.

## VI. CONCLUSION

There are many ways that a phishing attack that can be launched. Along with the technology development, the cybercriminals assets and methods are expanding too. Therefore, the matter is in our hands and thus there are a variety of tools, it is key for us to stay a step ahead and protect our systems and private information. Various types of phishing attacks have been studied, presented and identified. Another element of the success of protection is to anticipate, keep an eye open for future threats that are coming, keep our systems up to date and share the information.

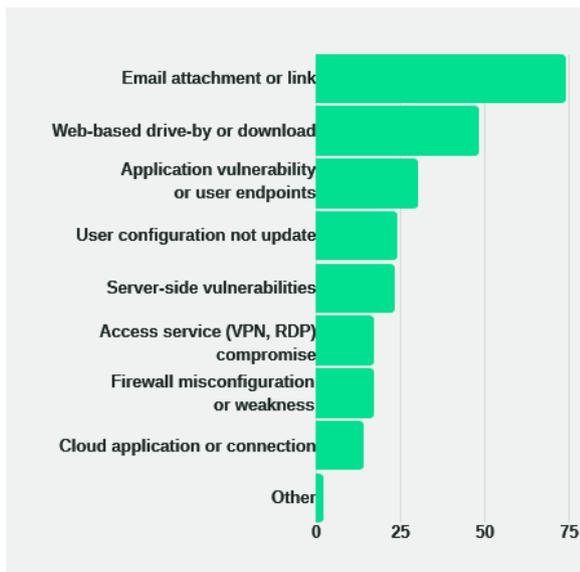

*Figure 6 – Top directions of entry [6]*

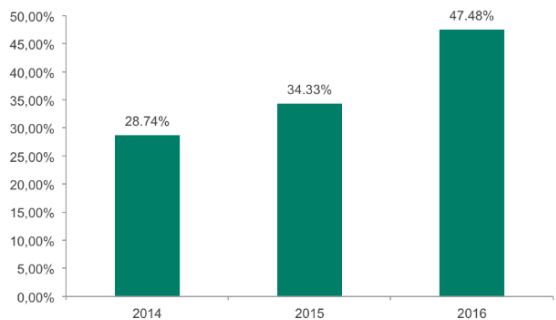

*Figure 7 – Percentage of financial phishing 2014-2016 [21]*

It is recommended that we stay alerted when surfing on the internet and be careful when sharing information. The public institutions and governments are also corroborating different laws to ensure we as citizens, are protected as much as possible. Some of the changes are creating difficult times for companies but improving the overall security of data. General Data Protection Regulation (GDPR) released in April 2016 constituted a major step forward for the European Economic Area (EEA) and provided users full control and transparency over the data.

Phishing will continue to pose a threat even if the law enforcements are doing their best to protect users and victims of the attack. Compared to other disciplines, cybersecurity will always be on high demand and attacks will constitute a problem that can never be fully solved.

REFERENCES

[1] Jason I. Hong, "The State of Phishing Attacks" Communications of the ACM, vol. 55, pp. 74-81, January 2012.

[2] Kang L. C.,Kelvin S. C. Y., Choon L. T, "A survey of pshising attacks: Their types, vectors and technical approaches" Expert Systems with Applications , vol. 106. Elsevier, 2018, pp.1–20.

[3] I. Ghafir, V. Prenosil, M. Hammoudeh and U. Raza, "Malicious SSL Certificate Detection: A Step Towards Advanced Persistent Threat Defence," International Conference on Future Networks and Distributed Systems. Cambridge, United Kingdom, 2017.

[4] FBI News, Stories: "Business E-mail Compromise," in Cyber-Enabled Financial Fraud on the Rise Globally, 2017. Website avilable at: https://www.fbi.gov/news/stories/business-e-mail-compromise-on-the-rise (accessed on 1st December 2019).

[5] U. Raza, J. Lomax, I. Ghafir, R. Kharel and B. Whiteside, "An IoT and Business Processes Based Approach for the Monitoring and Control of High Value-Added Manufacturing Processes," International Conference on Future Networks and Distributed Systems. Cambridge, United Kingdom, 2017.

[6] I. Ghafir and V. Prenosil, "Advanced Persistent Threat and Spear Phishing Emails." International Conference Distance Learning, Simulation and Communication. Brno, Czech Republic, pp. 34-41, 2015.

[7] Arpan C., Prashand K., Dinesh K. Y., "Phishing attacks and its countermeasurs," International Journals of Advanced Research in Computer Science and Software Engineering, vol. 7, 2017, pp. 245-253.

[8] I. Ghafir and V. Prenosil, "Malicious File Hash Detection and Drive-by Download Attacks," International Conference on Computer and Communication Technologies, series Advances in Intelligent Systems and Computing. Hyderabad: Springer, vol. 379, pp. 661-669, 2016.

[9] Ahmed A., Lina Z. , "Phishing environments, techniques, and countermeasures: A survey," Computers & Security, vol. 68, Elsevier, 2017, pp.160–196.

[10] Bhavsar V, Kadlak A., and Sharma S. Oka, "Study of Phishing Attacks", International Journal of Computer Applications, vol. 182, 33, 2018, pp. 27–29.

[11] I. Ghafir, J. Svoboda, V. Prenosil, "A Survey on Botnet Command and Control Traffic Detection," International Journal of Advances in Computer Networks and Its Security (IJCNS), vol. 5(2), pp. 75-80, 2015.

[12] J. Svoboda, I. Ghafir, V. Prenosil, "Network Monitoring Approaches: An Overview," International Journal of Advances in Computer Networks and Its Security (IJCNS), vol. 5(2), pp. 88-93, 2015.

[13] Storagepipe - Online Backup and Disaster Recovery. (2016) Phishing - Why Malware Happens. [Video] https://www.youtube.com/watch?v=PdTfofLhlfg Accessed 29th November 2019.

[14] I. Ghafir and V. Prenosil. "Proposed Approach for Targeted Attacks Detection," Advanced Computer and Communication Engineering Technology, Lecture Notes in Electrical Engineering. Phuket: Springer International Publishing, vol. 362, pp. 73-80, 9, 2016.

[15] Att Internet Service, Resources: "Beware: Fake Hotel Booking Sites", 2015. Website avilable at: https://www.attinternetservice.com/resources/fake-hotel-sites/ (accessed on 5th December 2019)

[16] I. Ghafir and V. Prenosil, "DNS query failure and algorithmically generated domain-flux detection," International Conference on Frontiers of Communications, Networks and Applications. Kuala Lumpur, Malaysia, pp. 1-5, 2014.

[17] Government of Australia, Scam Types:"Car Crash Componsation Phone Scam". (2019). Avilalbe at: https://www.scamnet.wa.gov.au/scamnet/Scam_types-Attempts_to_gain_your_personal_information-Phishing-Car_Crash_Compensation_Phone_Scam.htm (accesed on 29th November 2019)

[18] I. Ghafir, V. Prenosil, and M. Hammoudeh, "Botnet Command and Control Traffic Detection Challenges: A Correlation-based Solution." International Journal of Advances in Computer Networks and Its Security (IJCNS), vol. 7(2), pp. 27-31, 2017.

[19] Lee N., SANS Institute, Information Security Reading Room: "2017 Threat Landscape Survey: Users on the Front Line" , 2019.

[20] I. Ghafir, M. Husak and V. Prenosil, "A Survey on Intrusion Detection and Prevention Systems," IEEE/UREL conference, Zvule, Czech Republic, pp. 10-14, 2014.

[21] Karspersky., Secure List, Financial cyberthreats in 2016: "Big fish bring big rewards for cybercriminals in 2016 but that doesn't mean small fish are safe" , 2017.